\documentclass[10pt, conference]{IEEEtran}
\IEEEoverridecommandlockouts
\usepackage{cite}
\usepackage{amsmath,amssymb,amsfonts}
\usepackage{algorithmic}
\usepackage{graphicx}
\usepackage{textcomp}
\usepackage{xcolor}
\usepackage{enumitem}
\def\BibTeX{{\rm B\kern-.05em{\sc i\kern-.025em b}\kern-.08em
    T\kern-.1667em\lower.7ex\hbox{E}\kern-.125emX}}

\newtheorem{lemma}{\bf Lemma}
\newtheorem{proposition}{\bf Proposition}

\makeatletter
\newcommand*{\rom}[1]{\expandafter\@slowromancap\romannumeral #1@}
\makeatother

\begin{document}

\title{Bidirectional Information Flow and the Roles of Privacy Masks in Cloud-Based Control}

\author{\IEEEauthorblockN{ Ali Reza Pedram}
\IEEEauthorblockA{
\textit{University of Texas at Austin}\\
apedram@utexas.edu}
\and
\IEEEauthorblockN{Takashi Tanaka}
\IEEEauthorblockA{
\textit{University of Texas at Austin}\\
ttanaka@utexas.edu}
\and
\IEEEauthorblockN{Matthew Hale}
\IEEEauthorblockA{
\textit{University of Florida}\\
matthewhale@ufl.edu}
}

\maketitle

\begin{abstract}
We consider a cloud-based control architecture for a linear plant with Gaussian process noise, where the state of the plant contains a client's sensitive information. We assume that the cloud tries to estimate the state while executing a designated control algorithm. 
The mutual information between the
client's actual 
state and the cloud's estimate is adopted as a measure of privacy loss. 
We discuss the necessity of uplink and downlink privacy masks.
After observing that privacy is not necessarily a monotone function of the noise levels of privacy masks, we discuss the joint design procedure for uplink and downlink privacy masks. Finally, the trade-off between privacy and control performance is explored.   
\end{abstract}


\section{Introduction}
\label{secintro}
Cloud-based control is a popular platform in industrial automation in recent years. It has a number of advantages over conventional on-site controllers in terms of installation costs, flexibility, and computational resources. In a cloud-based control architecture, it is necessary for clients to share operational data of a local plant with the cloud in real time. Since operational data can contain clients' sensitive information, privacy concerns are inherent 
in cloud-based control. 

Privacy has been studied in recent years in a broad range of academic disciplines including control, information theory, and computer science. 
Formal concepts of privacy include $k$-anonymity, $t$-closeness, and $\ell$-diversity. The recent concept of differential privacy \cite{dwork2008differential} has been used in both the 
research community and industry.
Information-theoretic privacy \cite{sankar2013utility,Sudan} has also been investigated. 
Some of these privacy notions have already been applied to control-related problems. Differential privacy, in particular, has been applied to various  estimation and control problems  \cite{ny+14,wang+14,hale2015differentially,huang+12,sandberg+15,mo+16,nozari+16,cortes2016differential}. 
Information-theoretic privacy was also proposed in \cite{tanaka2017directed}, where a rate-distortion approach similar to \cite{sankar2013utility} was generalized to feedback control systems. 

While these results demonstrate the effectiveness of existing privacy notions in feedback control systems, it is important to be mindful that there are key differences between privacy problems for static (e.g., databases) and dynamic (e.g., feedback control) applications. The key feature of cloud-based control architectures, which distinguishes them from static databases, is the existence of a bidirectional information flow between the cloud and the client.
The purpose of this paper is to identify and address the 
privacy consequences of this fact, which 
has received comparatively little attention in
the existing privacy literature for feedback control systems. 

In this paper, we consider a cloud-based control architecture equipped with an \emph{uplink privacy mask} (perturbing the client-to-cloud information flow) and a \emph{downlink privacy mask} (perturbing the cloud-to-client information flow). Following \cite{sankar2013utility}, we adopt mutual information as a measure of privacy loss. Our problem setup is minimal, yet effective to draw the following conclusions:
\begin{itemize}
\item[(A)] Both an uplink privacy mask and a downlink privacy mask are necessary to enhance privacy, especially when the process noise of the plant is absent or small. Intuitively, this is because the absence of a downlink privacy mask allows the cloud to estimate the client's state solely based on control commands it sends. 
In this way, the cloud can compromise the client's privacy, even without ever receiving any information from the client. 
\item[(B)] Unlike applications in which only the uplink information flow exists, privacy is not necessarily a monotone function of the noise level of the uplink privacy mask. This fact can be understood by invoking the computation of the capacity of channels with feedback. Consequently, a joint design of uplink and downlink privacy masks is needed, which is demonstrated in this paper. 
\end{itemize}
\textbf{Notation}: Random variables are denoted by upper-case symbols. The notation $X^t=(X_0, X_1, ... , X_t)$ is used to denote a sequence.  If $X^T$ and $Y^T$ are random processes, the (forward) directed information \cite{massey1990causality} is defined as
\begin{equation}
I(X^T\rightarrow Y^T)\triangleq \sum\nolimits_{t=0}^T I(X^t; Y_t\mid Y^{t-1}).
\end{equation}
The (backward) directed information is denoted by
\begin{equation}
I(X^T\leftarrow Y^T)\triangleq I(0*X^{T-1}\rightarrow Y^T),
\end{equation}
where $0*X^{T-1}=(0, X_0, X_1, ... , X_{T-1})$. The natural logarithm is used throughout the paper. 

\section{Problem formulation}

\begin{figure}[t]
    \centering
    \includegraphics[width=\columnwidth]{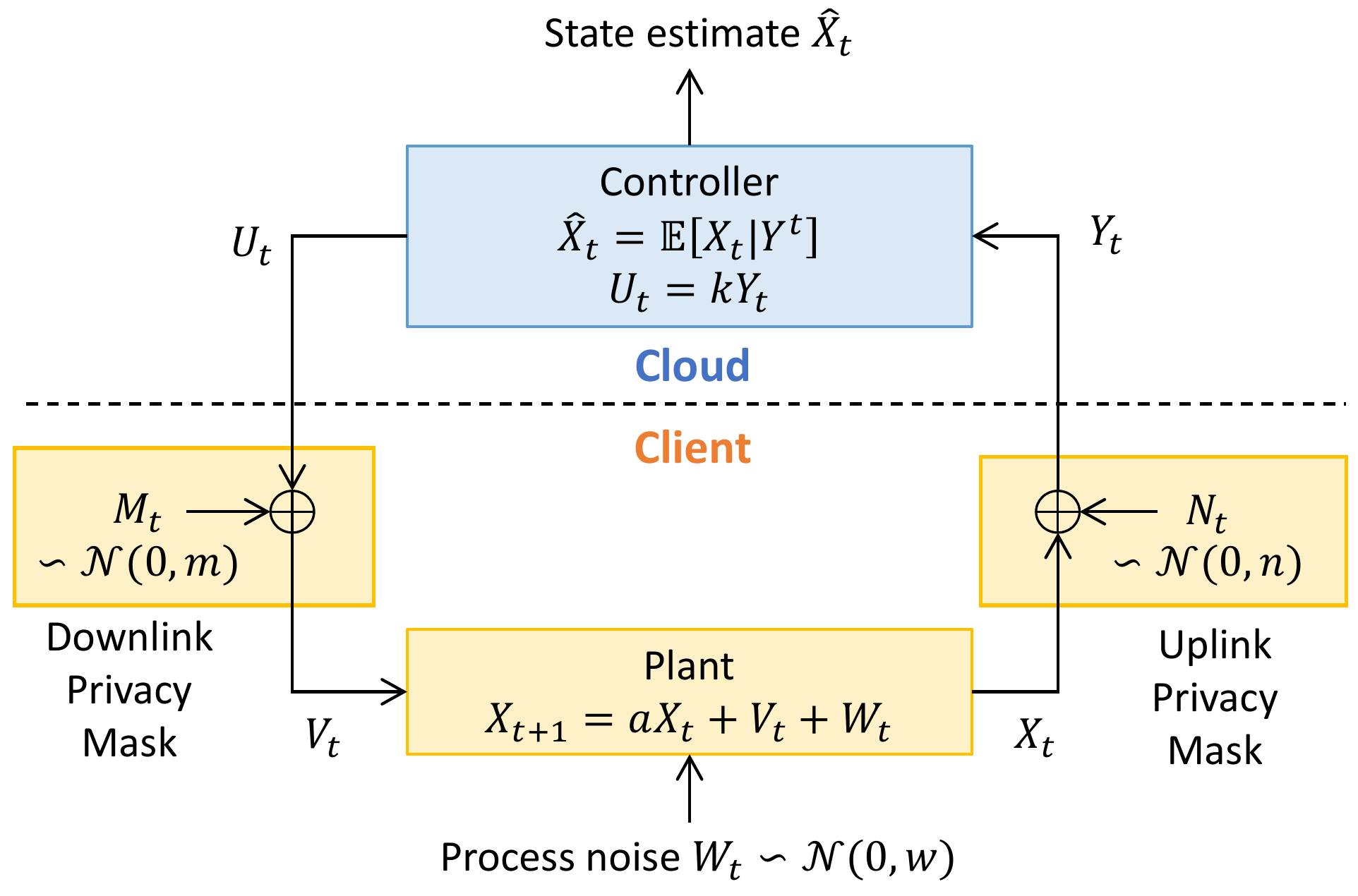}
    \caption{Cloud-based control architecture considered in this paper.}
    \label{fig:cloudcontrol}
\end{figure}

In this paper, we consider the cloud-based control architecture shown in Fig.~\ref{fig:cloudcontrol}. We assume there are two parties
in this configuration, 
namely the cloud operator and the client. The client owns a local plant, whose dynamics is described by a linear dynamical system with Gaussian process noise. The system parameters $a$ and $w$ are assumed to be known to both parties. We consider the situation in which the local plant is controlled by the cloud, but, in the interest of privacy, the client does not wish a full disclosure of the state of the plant, $X_t$. We assume that the client adopts a bidirectional privacy masking mechanism comprised of \emph{uplink} and \emph{downlink} privacy masks, as shown in Fig.~\ref{fig:cloudcontrol}. Both \emph{uplink} and \emph{downlink} privacy masks are simple additive white Gaussian noise (AWGN) channels, where noise intensities $m$ and $n$ are parameters to be designed. The cloud observes a perturbed version $Y_t=X_t+N_t$ of the state, which is multiplied by a feedback control gain $k\neq 0$ to compute the raw control input $U_t=kY_t$. Instead of applying the raw control input $U_t$, the client applies a perturbed version of the control input, $V_t=U_t+M_t$, where the realization of the random variable $M_t$ is not known to the cloud. For simplicity, we assume that all random variables are scalar-valued. We also assume that $k$ is a pre-designed feedback gain, which is treated as a constant in this paper.

To analyze the fundamental limitations of the cloud's capability to estimate $X_t$, we assume that the parameters $m$ and $n$ are available to the cloud, based on which the estimate $\hat{X}_t$ is computed by Kalman filtering. Due to the invertibility of the Kalman filter, the following identities are easy to verify:
\begin{align}
&I(X^T\rightarrow \hat{X}^T)=I(X^T\rightarrow Y^T)=I(X^T\rightarrow U^T) \\
&I(X^T\leftarrow \hat{X}^T)=I(X^T\leftarrow  Y^T)=I(X^T\leftarrow  U^T).
\end{align} 
Following \cite{sankar2013utility}, this paper adopts the mutual information $I(X^T; \hat{X}^T)$ as a privacy metric. The following identity, known as the \emph{conservation of information} \cite{massey2005conservation}, is useful in the developments below:
\begin{equation}
      I(X^T;\hat{X}^T)=I(X^T\rightarrow \hat{X}^T)+I(X^T\leftarrow \hat{X}^T).
  \end{equation}
This identity articulates the fact that it is not only the uplink information flow, but also the downlink information flow that impact the confidentiality of cloud-based control systems. Minimizing
mutual information should therefore account for information
flow in both directions.

\section{Main Results}

 \begin{figure}[t]
    \centering
    \includegraphics[width=\columnwidth]{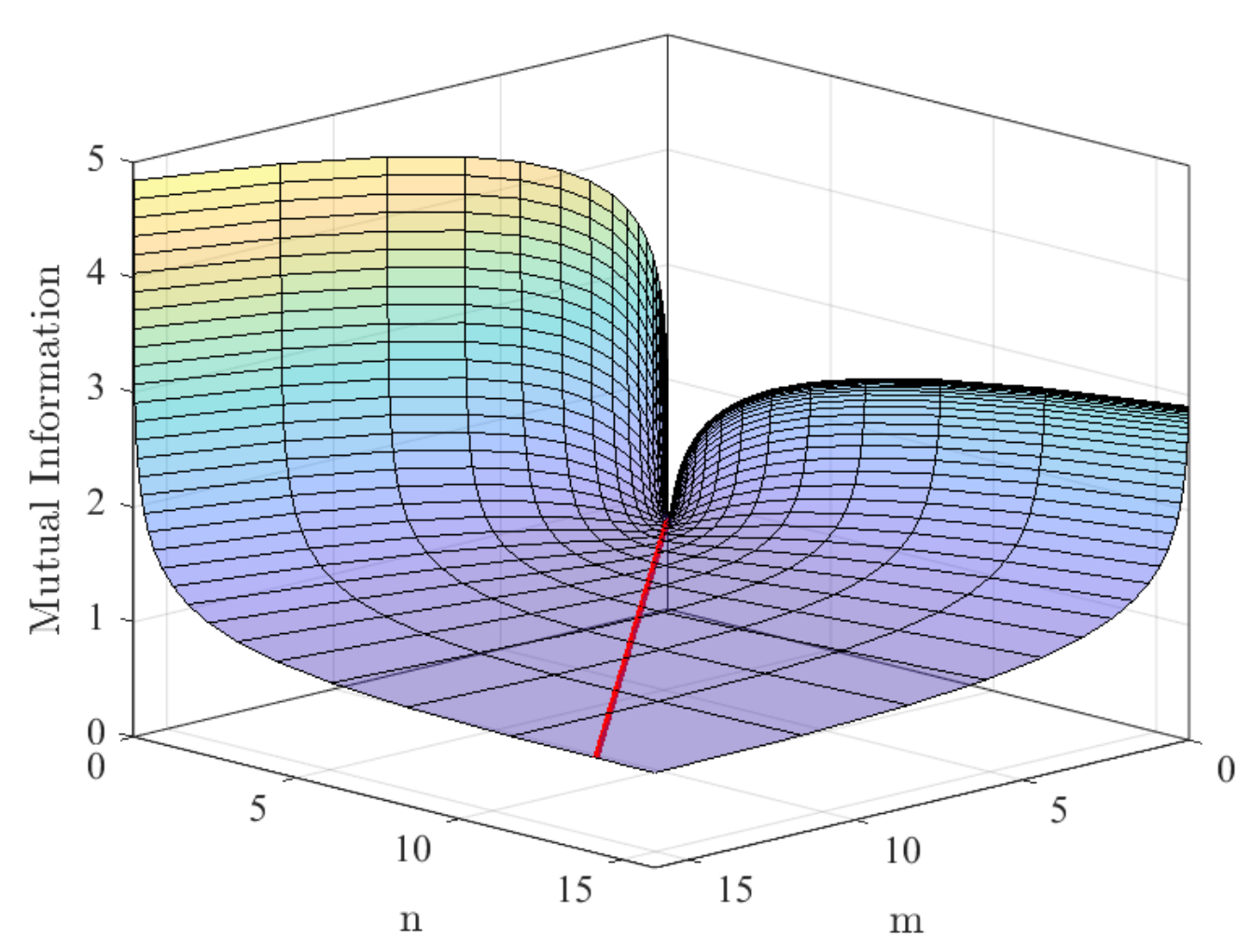}
    \caption{Mutual Information for different values of $m$ and $n$ and for $a=1$, $k=-1$, and $w=0.05$.}
    \label{fig:3d}
\end{figure}

\subsection{Computation of performance metrics}
To analyze performance from both information and control perspectives, we explicitly compute the following metrics:
\begin{align}
&I_\infty(X\rightarrow \hat{X}) \triangleq \limsup_{T\rightarrow \infty} \frac{1}{T+1} I(X^T\rightarrow \hat{X}^T)\\ 
&I_\infty(X\leftarrow \hat{X}) \triangleq \limsup_{T\rightarrow \infty} \frac{1}{T+1} I(X^T\leftarrow \hat{X}^T)\\ 
&I_\infty(X; \hat{X}) \triangleq \limsup_{T\rightarrow \infty} \frac{1}{T+1} I(X^T; \hat{X}^T)\\ 
&C_\infty(X, U)\triangleq \limsup_{T\rightarrow \infty} \frac{1}{T+1} \sum_{t=0}^T \mathbb{E}[qX_t^2+rU_t^2].
\end{align}
In what follows, we regard $I_\infty(X; \hat{X})$ as the privacy loss
and $C_\infty(X, U)$ as the control cost. In general, it is desirable to keep both of these quantities small. We next derive exact expressions for them in terms of system parameters.
\begin{lemma}
\label{lem1}
The mutual information and the $H_2$ control performance are 
\begin{align} 
I_\infty(X; \hat{X})=&\underbrace{\frac{1}{2}\log\left(1+\frac{\Sigma}{n}\right) }_{I_\infty(X\rightarrow \hat{X})}+\underbrace{\frac{1}{2}\log\left(1+\frac{k^2n}{m+w}\right)}_{I_\infty(X\leftarrow \hat{X})}  \label{eqmi_inf} \\
C_\infty(X, U)=&\frac{(q+rk^2)(m+k^2 n+w)}{1-(a+k)^2}+rk^2n, \label{eqc_inf}
\end{align}
where $\Sigma$ is the positive solution to the algebraic Riccati equation (ARE)
\begin{equation}
\label{ARE}
    \left(\frac{a^2n}{\Sigma+n}-1\right)\Sigma+m+w=0.
\end{equation}
\end{lemma}

Fig.~\ref{fig:3d} shows the values of mutual information $I_\infty(X; \hat{X})$ as a function of $m$ and $n$. We note that it is neither convex nor concave in $(m,n)$.

\subsection{Necessity of privacy masks}
The necessity of privacy masks can be understood by analyzing  $I_\infty(X; \hat{X})$. 
\begin{proposition}
\label{prop1}
For arbitrary $w\geq0$ and $m\geq0$ such that $w+m>0$, we have
$\lim_{n\searrow 0} I_\infty(X; \hat{X})=+\infty$.
\end{proposition}

This result shows that the lack of uplink privacy mask (cloud observes unperturbed data) implies unbounded privacy loss, indicating its necessity for privacy protection. 
\begin{proposition}
\label{prop2}
If $w=0$, for arbitrary $n\geq0$, we have
$\lim_{m\searrow 0} I_\infty(X; \hat{X})=+\infty$. 
\end{proposition}

This result indicates that the downlink privacy mask is also necessary when $w=0$. To see why, notice that when $w=0$, the lack of downlink privacy mask ($m=0$) makes the plant a deterministic dynamical system from the cloud's perspective. Assuming that the initial state $X_0$ is known, the cloud operator is able to compute $X_t$ by ``simulating'' the dynamics based on the knowledge of the control sequence $U^t$ alone.  We emphasize that such a privacy attack through the downlink channel is possible even when the
client shares no information over time. 

Proposition~\ref{prop2} does not apply when $w\neq 0$.
However, even in such cases, the downlink privacy filter can
provide robustness against modeling errors. We will revisit this point in Section~\ref{sectradeoff}.

\subsection{Connection to the capacity of channels with feedback}

As Fig.~\ref{fig:3d} shows, for a fixed value of $m$, the privacy loss \eqref{eqmi_inf} is \emph{not} a monotonically decreasing function of the noise level $n$ of the uplink privacy filter. This is a unique characteristic of systems with feedback, and shows a stark contrast to the privacy theory for static databases, 
where larger perturbations provide stronger privacy protections. 
A closer look at \eqref{eqmi_inf} reveals that, for fixed values of $(m, w, a, k)$, the uplink directed information $I(X\rightarrow \hat{X})$ is indeed a convex and monotonically non-increasing function of $n$. A related observation has already been made in \cite{tanaka2017directed}. However, the downlink directed information $I_\infty(X\leftarrow \hat{X})$ is a concave and monotonically increasing function of $n$. To understand the latter fact, it is instrumental to invoke the computation of channel capacity under  noiseless feedback \cite{cover1989gaussian,permuter2009finite,kourtellaris2018information,pedram2018some}.

\begin{figure}[t]
    \centering
    \includegraphics[width=0.8\columnwidth]{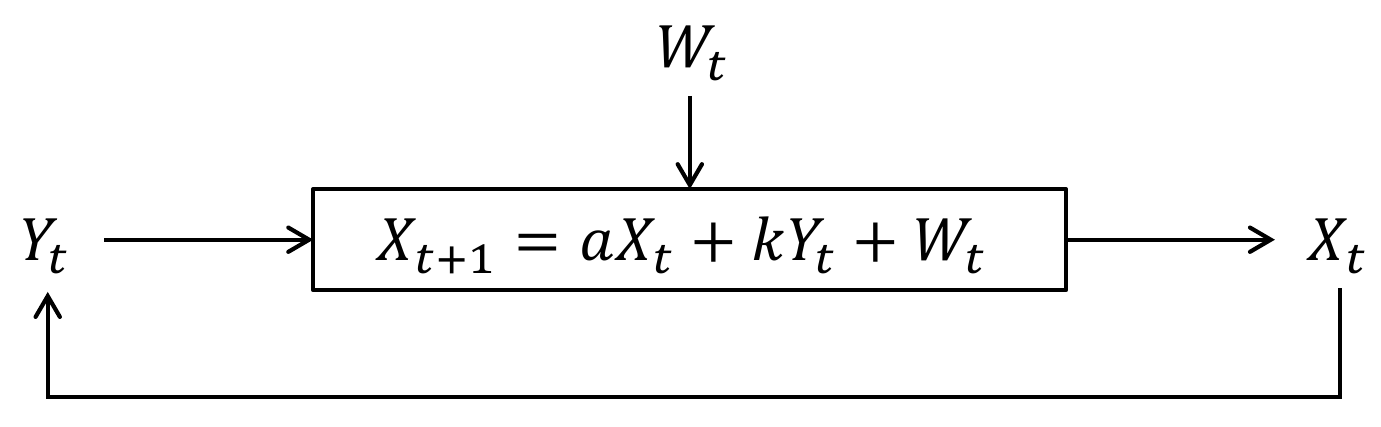}
    \caption{When $m=0$, Fig.~\ref{fig:cloudcontrol} can be viewed as a communication channel with memory with noiseless feedback.}
    \label{fig:feedbackchannel}
\end{figure}

Fig.~\ref{fig:feedbackchannel} shows an equivalent diagram to Fig.~\ref{fig:cloudcontrol} when $m=0$. This diagram shows a connection between Fig.~\ref{fig:cloudcontrol} and the computation of the capacity of a channel with feedback. It is known that the capacity of a channel with noiseless feedback coincides with the supremum of $I_\infty(X\leftarrow Y)$ over the distribution of inputs, subject to channel input constraints \cite{permuter2009finite}. 
In Fig.~\ref{fig:cloudcontrol} with $m=0$, the channel input distribution is restricted to the form $Y_t=X_t+N_t$, with $N_t\sim\mathcal{N}(0,n)$. Since  the channel input power $\mathbb{E}[Y_t^2]$ monotonically increases with $n$, the capacity also increases with $n$.
The second term of the equation \eqref{eqmi_inf} confirms this intuition.

\subsection{Joint uplink and downlink mask designs}

\begin{figure}[t]
    \centering
    \includegraphics[width=\columnwidth]{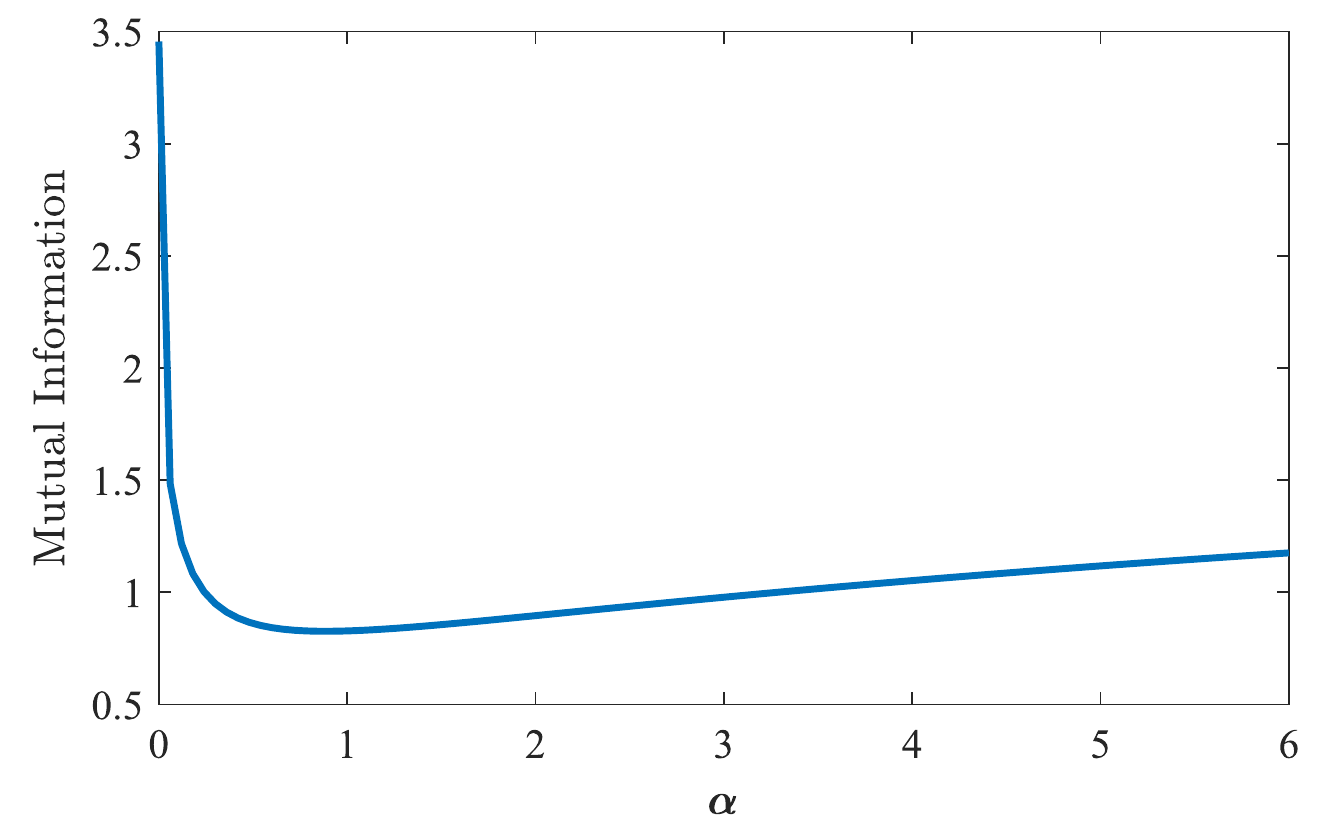}
    \caption{Mutual Information $I_\infty(X^T;\hat{X}^T)$ as a function of $\alpha$ for $a=1$, $k=-1$, and $w=0.05$.}
    \label{fig:m_fixed}
\end{figure}

In this section, we find the optimal values of $m$ and $n$ to have  minimum privacy loss.
\begin{proposition}
\label{prop3}
The mutual information \eqref{eqmi_inf} depends only on the Noise-to Noise (NNR) ratio $\alpha := \frac{n}{m+w}$. In particular, 
\begin{align}
    I_\infty(X; \hat{X})&=
    \frac{1}{2}\log \left(\frac{(a^2-1+\frac{1}{\alpha})+\sqrt{(a^2-1+\frac{1}{\alpha})^2+\frac{4}{\alpha}}}{2}\right) \nonumber \\
    &+\frac{1}{2}\log\left(1+k^2\alpha\right). \label{eqI_alpha}
\end{align}
\end{proposition}
Proposition~\ref{prop3} indicates that the level of privacy is determined by the ratio of $n$ and $m+w$ and not by the magnitude of noises. The mutual information \eqref{eqI_alpha} as a function of $\alpha$ is depicted in Fig.~\ref{fig:m_fixed}. This function is neither convex nor concave. Nevertheless, we have the following result:
\begin{proposition}
\label{prop4}
Mutual information \eqref{eqI_alpha} is uniquely minimized by the unique positive solution $\alpha^*$ to the quartic equation
\begin{align}
    (a^2-1)^2 \alpha^4+2(a^2+1)\alpha^3-\frac{2}{k^2}\alpha-\frac{1}{k^4}=0.
\end{align}
\end{proposition}

The red line on Fig.~\ref{fig:3d} shows the line $n=\alpha^*(m+w)$. As 
can be inferred, $I_\infty(X;\hat{X})$ is constant along this line. 

\subsection{Privacy-utility trade-off}
\label{sectradeoff}
The privacy masks also alter the control performance $C_\infty(X,U)$. To analyze the privacy-utility trade-off, we now consider the following multi-objective optimization problem
 \begin{align}
 \label{opt_ci}
     \min_{m,n} I_\infty (X;\hat{X})+\lambda C_\infty(X,U)
 \end{align}
where $\lambda>0$ is a positive weight.
We observed in Lemma~\ref{lem1} that the control cost is affine in $m$ and $n$. 
Thus, using~$n = \alpha(m+w)$, we can express $C_{\infty}$ as an affine function of $\alpha$, $m$, and $\alpha m$, which is denoted by $C_\infty (\alpha,\alpha m,m)$ with a slight abuse of notation.
Since Proposition~\ref{prop3} shows that $I_\infty (X;\hat{X})$ is only a function of $\alpha$ (denoted by $I_\infty (\alpha)$), \eqref{opt_ci} can be rewritten in terms of $\alpha$ and $m$ as $\min_{\alpha,m} \big[ I_\infty (\alpha)+\lambda \  C_\infty (\alpha,\alpha m,m)\big]$. Assuming that the closed-loop system is stable (i.e., $|a+k|<1$), the affine function $C_\infty (\alpha,\alpha m,m)$ is increasing in its individual arguments. Hence, we can proceed as
\begin{subequations}
\label{opt_ci2}
\begin{align}
     &\min_{\alpha,m} \big[ I_\infty (\alpha)+\lambda \  C_\infty (\alpha,\alpha m,m)\big]\\ \label{opt_b}
      &=\min_{\alpha} \big[ I_\infty (\alpha)+\lambda \min_{m} [C_\infty (\alpha,\alpha m,m)]\big]\\\label{opt_c}
      &=\min_{\alpha} \big[ I_\infty (\alpha)+\lambda\  C_\infty (\alpha,0,0)\big], 
\end{align}
\end{subequations}
where $C_\infty(\alpha,0,0)$ is an increasing affine function in $\alpha$.
This observation indicates that (i) the privacy-utility trade-off can be performed purely in terms of the NNR $\alpha=\frac{n}{m+w}$, and that (ii) as far as the NNR $\alpha$ is chosen to be the minimizer of \eqref{opt_c}, selecting $m=0$ (no downlink mask) is mathematically optimal. However, (ii) must be taken with two caveats. First, as observed in Proposition~\ref{prop2}, simultaneous $m=0$ and $w=0$ result in an ill-defined $\alpha$ and an unbounded privacy loss. Second, even if $w>0$, the task of selecting the optimal $n$ such that $\alpha=\frac{n}{m+w}$ minimizes \eqref{opt_c} becomes sensitive to  modeling error of $w$ if $m=0$. Thus, the downlink mask has the practical
benefit of enhancing robustness against modeling errors of $w$.

\section{Proofs}
\subsection{Proof of Lemma~\ref{lem1}}
We consider the case of a scalar-valued LTI system in finite time horizon. The closed-loop dynamics are
\begin{align}
    &X_t=a X_{t-1}+U_{t-1}+ M_{t-1}+W_t\\
    &Y_t=X_t+N_t\\
    &U_t=kY_t.
\end{align}
 $X_t$ can be estimated by
$\hat{X}_t=\mathbb{E}[X_t \mid Y^t,U^{t-1}]$ with the error covariance  $\Sigma_{t}=\mathbb{E}[(X_t-\hat{X}_t)(X_t-\hat{X}_t)^\intercal\mid Y^t,U^{t-1}]$ that can be calculated iteratively by Kalman 
filtering via
\begin{subequations}
\label{KF}
\begin{align}
    &\hat{X}_{t|t-1}=a \hat{X}_{t-1}+ U_{t-1} \quad  \quad  \hat{X}_0=0\\
    &\Sigma_{t|t-1}=a^2 \Sigma_{t-1}+ m+w \quad \Sigma_{0}=0\\ \label{Ric1}
    &l_t=\Sigma_{t|t-1} /(\Sigma_{t|t-1}+n)\\
    & \hat{X}_{t}=\hat{X}_{t|t-1}+l_t(Y_t-\hat{X}_{t|t-1})\\ \label{Ric2}
    &\Sigma_{t}=(1-l_t)^2 \Sigma_{t|t-1}+l_t^2 n.
\end{align}
\end{subequations}
It follows from (\ref{KF}) that
  \begin{equation}
  \label{expan}
      \begin{split}
          &\hat{X}_t=\hat{X}_{t|t-1}+l_t(X_t+N_t-\hat{X}_{t|t-1})\\
          &=(1-l_t)a\hat{X}_{t-1}+(1-l_t)U_{t-1}+l_t X_t+l_t N_t.
      \end{split}
  \end{equation}
The uplink information flow $I(X^T\rightarrow \hat{X}^T)$ is computed as
\begin{equation}
     \begin{split}
         &I(X^T\rightarrow \hat{X}^T)=\sum\nolimits_{t=0}^T I(X^t;\hat{X}_t |\hat{X}^{t-1})=\\
         &\sum\nolimits_{t=1}^T \Big[  \underbrace{I(X^{t-1};\hat{X}_t |\hat{X}^{t-1}, X_t)}_{=(A)}+\underbrace{I(X_t;\hat{X}_t |\hat{X}^{t-1})}_{=(B)} \Big].
     \end{split}
 \end{equation}
Considering \eqref{expan} and the fact that $U_{t-1}=kY_{t-1}$ is a deterministic function of $\hat{X}^{t-1}$, we have:
 \[(A)=I(X^{t-1};\hat{X}_t |\hat{X}^{t-1}, X_t)=I(X^{t-1};l_t N_t|\hat{X}^{t-1},X_t)=0\]
 since $N_t$ is independent of $X^{t-1}$ given $(\hat{X}^{t-1},X_t)$. We next focus on $(B)$, again by considering (\ref{expan}):
 \begin{equation}
      \begin{split}
      &(B)=I(X_{t};\hat{X}_t |\hat{X}^{t-1})\\
      &=I(X_{t};(1-l_t)U_{t-1}+l_t X_t+l_tN_t|\hat{X}^{t-1})\\
      &=I(X_t;l_tX_t+l_tN_t|\hat{X}^{t-1})\\
      &=h(l_tX_t+l_tN_t|\hat{X}^{t-1})-h(l_tX_t+l_tN_t|\hat{X}^{t-1},X_t)\\
      &=h(l_tX_t+l_tN_t|\hat{X}^{t-1})-\frac{1}{2}\log(l_t^2 n). 
   \end{split}
  \end{equation}
The first term in the last line can be made more explicit as
 \begin{subequations}
 \label{rom3}
     \begin{align}
     \nonumber
         &h(l_t(aX_{t-1}+U_{t-1}+M_{t-1}+W_t+N_t)|\hat{X}^{t-1})\\
         &=h(l_t(aX_{t-1}+M_{t-1}+W_t+N_t)|\hat{X}^{t-1})\\\label{c}
         &=h(l_t(ae_{t-1}+M_{t-1}+W_t+N_t)|\hat{X}^{t-1})\\
         &=\frac{1}{2}\log(l_t^2(a^2\Sigma_{t-1}+m+w+n))\\
         &=\frac{1}{2}\log(l_t^2(\Sigma_{t|t-1}+n)).
     \end{align}
 \end{subequations}
 In (\ref{c}), we used the fact the innovation $e_{t-1}=X_{t-1}-\hat{X}_{t-1}$ is independent of the previous measurement (and of $\hat{X}^{t-1}$). Therefore, the uplink directed information is:
 \begin{align*}
     I(X^T\rightarrow \hat{X}^T)&=\sum\nolimits_{t=1}^T \Big[\frac{1}{2}\log(l_t^2(\Sigma_{t|t-1}+n))-\frac{1}{2}\log(l_t^2 n)\Big]\\
     &=\sum\nolimits_{t=1}^T  \frac{1}{2}\log\left(1+\frac{\Sigma_{t|t-1}}{n}\right).
 \end{align*}
The backward directed information is computed as
  \begin{equation}
  \nonumber
      \begin{split}
         &I(0*\hat{X}^{T-1}\rightarrow X^T)=I(0;X_1|X_0)+\sum_{t=1}^T I(\hat{X}^{t-1};X_t|X^{t-1})\\
         &= \sum\nolimits_{t=1}^T \Big[h(X_t|X^{t-1})-h(X_t|\hat{X}^{t-1},X^{t-1})\Big]\\
         &= \sum\nolimits_{t=1}^T \Big[h(kN_{t-1}+M_{t-1}+W_{t-1})-h(M_{t-1}+W_{t-1})\Big]\\
        &= \sum\nolimits_{t=1}^T \Big[\frac{1}{2}\log(k^2n +m+w)-\frac{1}{2}\log(m+w)\Big]\\
        & =\sum\nolimits_{t=1}^T \Big[\frac{1}{2}\log\left(1+\frac{k^2n}{m+w}\right)\Big].
      \end{split}
  \end{equation}
Due to the observability of the system, the limit $\Sigma= \lim_{t\rightarrow \infty} \Sigma_{t|t-1}$ exists for the Riccati recursion defined by (\ref{Ric1}) and (\ref{Ric2}), which coincide with the unique positive solution of ARE (\ref{ARE}).
 
 To compute the control cost $C_\infty(X, U)$, consider the predicted error covariance $P_t=\mathbb{E}[ X_t^2]$ which can be computed from the recursion
 \begin{align}
  \label{pri}
    P_t=(a+k)^2 P_{t-1}+m+k^2n+w  \quad P_0=0.
 \end{align}
The control cost is obtained as
 \begin{align*}
     &C_\infty(X, U)= \limsup_{T\rightarrow \infty} \frac{1}{T} \sum\nolimits_{t=1}^T \mathbb{E}[qX_t^2+rU_t^2]\\
     &=\limsup_{T\rightarrow \infty} \frac{1}{T} \sum\nolimits_{t=1}^T \mathbb{E}[(q+rk^2)X_t^2+rk^2N_t^2]\\
     &= \limsup_{T\rightarrow \infty} \frac{1}{T} \sum_{t=1}^T [ (q+rk^2)P_t +rk^2n]\!=\!(q+rk^2) P\!+rk^2n,
 \end{align*}
 where $P=\frac{m+k^2 n+w}{1-(a+k)^2}$ is the solution to algebraic Riccati equation induced by (\ref{pri}).

 \subsection{Proof of Propositions \ref{prop1} and \ref{prop2}}
From (\ref{ARE}), $\Sigma \rightarrow m+w$ as $n\rightarrow 0$. Thus, Proposition~\ref{prop1} is clear. When $w=0$,  the downlink information flow diverges for $m \rightarrow 0$ unless $m+w=0$ (Proposition \ref{prop2}). 
 
 \subsection{Proof of Proposition \ref{prop3}}
For uplink directed information, by direct substitution,
\begin{align}
\nonumber
   I_\infty(X\leftarrow \hat{X})=\frac{1}{2}\log\left(1+\frac{k^2n}{m+w}\right)= \frac{1}{2}\log\left(1+k^2\alpha\right).
\end{align}
To compute the downlink directed information, we can plug $m+w=\frac{n}{\alpha}$ in  
ARE (\ref{ARE}). Then (\ref{ARE}) reduces to
\begin{align}
\nonumber
    \left(\frac{\Sigma}{n}\right)^2-\left(a^2-1+\frac{1}{\alpha}\right)\frac{\Sigma}{n}-\frac{1}{\alpha}=0.
\end{align}
Substituting the positive solution in $I_\infty(X\rightarrow \hat{X})=\frac{1}{2}\log\left(1+\frac{\Sigma}{n}\right)$ completes the proof. 

\subsection{Proof of Proposition \ref{prop4}}

Equation \eqref{ARE} can be solved for $n$ as
$n=\frac{\Sigma(\Sigma-p)}{(a^2-1)\Sigma+p}$,
where $p:=m+w$ is defined for simplicity. Substituting this into (\ref{eqmi_inf}), we have
\begin{equation*}
I_\infty (X;\hat{X})\!=\!\frac{1}{2}\log\! \underbrace{ \left( \frac{a^2\Sigma}{\Sigma-p}\!+\!\frac{k^2}{p} \frac{\Sigma^2}{(a^2\!-\!1)\Sigma\!+p} \right)}_{=F}.
\end{equation*}
Since $F>0$, the minimizing 
$\Sigma$ is obtained by solving 
\begin{align}
    \label{diff}
    \frac{1}{a^2}\frac{\partial F}{\partial \Sigma}= \frac{-p}{(\Sigma-p)^2}+\frac{k^2}{p}\frac{(a^2-1)\Sigma^2+2p\Sigma}{((a^2-1)\Sigma+p)^2}=0.
\end{align}
Substituting $(a^2-1)\Sigma+p=\frac{\Sigma(\Sigma-p)}{n}$ (this identity follows from \eqref{ARE}) into the denominator of the second term in \eqref{diff}, we obtain 
\[
 \frac{1}{ (\Sigma-p)^2} \left(-p\Sigma+\frac{n^2k^2}{p}((a^2-1)\Sigma+2p)\right)=0
\]
from which $\Sigma$ is solved as $\Sigma=\frac{2pn^2}{(\frac{p^2}{k^2}-(a^2-1)n^2)}$.
Substituting this back into \eqref{ARE}, a quartic equation for $n$ is obtained as
\begin{align}
\label{optn}
    (a^2-1)^2n^4+2p(a^2+1)n^3-2\frac{p^3}{k^2}n-\frac{p^4}{k^4}=0.
\end{align}
We can rewrite (\ref{optn}) in terms of $\alpha=\frac{n}{p}$ as
\begin{align}
\label{optalpha}
F(\alpha)=(a^2-1)^2\alpha^4+2(a^2+1)\alpha^3\!-\!\frac{2}{k^2}\alpha-\frac{1}{k^4}=0.
\end{align}
Note that $F'(0)<0$, $\lim_{\alpha\rightarrow \infty} F'(\alpha)=\infty$, and $F'(\alpha)$ is strictly increasing. Thus, by the intermediate value theorem, there exists a unique $\hat{\alpha}>0$ such that $F'(\hat{\alpha})=0 $. Therefore, $F(\alpha)$ is strictly decreasing  for $\alpha \in [0,\hat{\alpha})$ and strictly increasing for $\alpha \in (\hat{\alpha},\infty)$. It follows from $F(0)<0$ that $F(\hat{\alpha})<0$. Since $\lim_{\alpha\rightarrow \infty} F(\alpha)=\infty$ and $F(\alpha)$ is strictly increasing on $(\hat{\alpha},\infty)$, again by the intermediate value theorem, $F(\alpha)=0$ necessarily has a unique positive solution $\alpha^*$.


\section{Future Work}

Future work includes the generalization of this paper's results to multi-dimensional control systems. It is also worth investigating whether the observations (A) and (B) in Section~\ref{secintro} hold for other privacy metrics, such as differential privacy.

\bibliography{references}

\begin{thebibliography}{10}
\providecommand{\url}[1]{#1}
\csname url@samestyle\endcsname
\providecommand{\newblock}{\relax}
\providecommand{\bibinfo}[2]{#2}
\providecommand{\BIBentrySTDinterwordspacing}{\spaceskip=0pt\relax}
\providecommand{\BIBentryALTinterwordstretchfactor}{4}
\providecommand{\BIBentryALTinterwordspacing}{\spaceskip=\fontdimen2\font plus
\BIBentryALTinterwordstretchfactor\fontdimen3\font minus
  \fontdimen4\font\relax}
\providecommand{\BIBforeignlanguage}[2]{{%
\expandafter\ifx\csname l@#1\endcsname\relax
\typeout{** WARNING: IEEEtran.bst: No hyphenation pattern has been}%
\typeout{** loaded for the language `#1'. Using the pattern for}%
\typeout{** the default language instead.}%
\else
\language=\csname l@#1\endcsname
\fi
#2}}
\providecommand{\BIBdecl}{\relax}
\BIBdecl

\bibitem{dwork2008differential}
C.~Dwork, ``Differential privacy: A survey of results,'' \emph{International
  Conference on Theory and Applications of Models of Computation}, pp. 1--19,
  2008.

\bibitem{sankar2013utility}
L.~Sankar, S.~R. Rajagopalan, and H.~V. Poor, ``Utility-privacy tradeoffs in
  databases: An information-theoretic approach,'' \emph{IEEE Transactions on
  Information Forensics and Security}, vol.~8, no.~6, pp. 838--852, 2013.

\bibitem{Sudan}
B.~Chor, O.~Goldreich, E.~Kushilevitz, and M.~Sudan, ``Private information
  retrieval,'' \emph{IEEE Symposium on Foundations of Computer Science}, 1995.

\bibitem{ny+14}
J.~L. Ny and G.~J. Pappas, ``Differentially private filtering,'' \emph{IEEE
  Transactions on Automatic Control}, vol.~59, no.~2, pp. 341--354, Feb 2014.

\bibitem{wang+14}
Y.~Wang, Z.~Huang, S.~Mitra, and G.~E. Dullerud, ``Entropy-minimizing mechanism
  for differential privacy of discrete-time linear feedback systems,''
  \emph{The 53rd IEEE Conference on Decision and Control (CDC)}, 2014.

\bibitem{hale2015differentially}
M.~Hale and M.~Egerstedt, ``Differentially private cloud-based multi-agent
  optimization with constraints,'' \emph{American Control Conference (ACC)},
  pp. 1235--1240, 2015.

\bibitem{huang+12}
Z.~Huang, S.~Mitra, and G.~Dullerud, ``Differentially private iterative
  synchronous consensus,'' \emph{Proceedings of the ACM Workshop on Privacy in
  the Electronic Society}, 2012.

\bibitem{sandberg+15}
H.~Sandberg, G.~D{\'a}n, and R.~Thobaben, ``Differentially private state
  estimation in distribution networks with smart meters,'' \emph{The 54th IEEE
  Conference on Decision and Control (CDC)}, 2015.

\bibitem{mo+16}
Y.~Mo and R.~M. Murray, ``Privacy preserving average consensus,'' \emph{IEEE
  Transactions on Automatic Control}, vol.~62, no.~2, pp. 753--765, 2017.

\bibitem{nozari+16}
E.~Nozari, P.~Tallapragada, and J.~Cort{\'e}s, ``Differentially private
  distributed convex optimization via objective perturbation,'' \emph{2016
  American Control Conference (ACC)}, 2016.

\bibitem{cortes2016differential}
J.~Cort{\'e}s, G.~E. Dullerud, S.~Han, J.~Le~Ny, S.~Mitra, and G.~J. Pappas,
  ``Differential privacy in control and network systems,'' \emph{Decision and
  Control (CDC), 2016 IEEE 55th Conference on}, pp. 4252--4272, 2016.

\bibitem{tanaka2017directed}
T.~Tanaka, M.~Skoglund, H.~Sandberg, and K.~H. Johansson, ``Directed
  information and privacy loss in cloud-based control,'' \emph{American Control
  Conference (ACC), 2017}, pp. 1666--1672, 2017.

\bibitem{massey1990causality}
J.~Massey, ``Causality, feedback and directed information,''
  \emph{International Symposium on Information Theory and Its Applications
  (ISITA)}, 1990.

\bibitem{massey2005conservation}
J.~L. Massey and P.~C. Massey, ``Conservation of mutual and directed
  information,'' \emph{International Symposium on Information Theory (ISIT)},
  pp. 157--158, 2005.

\bibitem{cover1989gaussian}
T.~M. Cover and S.~Pombra, ``Gaussian feedback capacity,'' \emph{IEEE
  Transactions on Information Theory}, vol.~35, no.~1, pp. 37--43, 1989.

\bibitem{permuter2009finite}
H.~H. Permuter, T.~Weissman, and A.~J. Goldsmith, ``Finite state channels with
  time-invariant deterministic feedback,'' \emph{IEEE Transactions on
  Information Theory}, vol.~55, no.~2, pp. 644--662, 2009.

\bibitem{kourtellaris2018information}
C.~K. Kourtellaris and C.~D. Charalambous, ``Information structures for
  feedback capacity of channels with memory and transmission cost: Stochastic
  optimal control and variational equalities,'' \emph{IEEE Transactions on
  Information Theory}, vol.~64, no.~7, pp. 4962--4992, 2018.

\bibitem{pedram2018some}
A.~Pedram and T.~Tanaka, ``Some results on the computation of feedback capacity
  of {G}aussian channels with memory,'' \emph{56th Annual Allerton Conference
  on Communication, Control, and Computing}, 2018.

\end{thebibliography}
\bibliographystyle{IEEEtran}

\end{document}